# Impact of Correlated Failures in 5G Dual Connectivity Architectures for URLLC Applications

Milad Ganjalizadeh*[†], Piergiuseppe Di Marco[‡], Jonas Kronander*, Joachim Sachs*, and Marina Petrova[†]

*Ericsson Research, Ericsson AB, Stockholm, Sweden
Email: {milad.ganjalizadeh, jonas.kronander, joachim.sachs}@ericsson.com

[†]School of Electrical Engineering and Computer Science, KTH Royal Institute of Technology, Sweden
Email: petrovam@kth.se

[‡]Department of Information Engineering, Computer Science and Mathematics, University of L'Aquila, Italy
Email: piergiuseppe.dimarco@univaq.it

*Abstract*—Achieving end-to-end ultra-reliability and resiliency in mission critical communications is a major challenge for future wireless networks. Dual connectivity has been proposed by 3GPP as one of the viable solutions to fulfill the reliability requirements. However, the potential correlation in failures occurring over different wireless links is commonly neglected in current network design approaches. In this paper, we investigate the impact of realistic correlation among different wireless links on end-to-end reliability for two selected architectures from 3GPP. In ultra-reliable use-cases, we show that even small values of correlation can increase the end-to-end error rate by orders of magnitude. This may suggest alternative feasible architecture designs and paves the way towards serving ultra-reliable communications in 5G networks.

## I. INTRODUCTION

Ultra-reliable low-latency communication (URLLC) is characterized by stringent reliability and delay requirements needed by many types of new applications in the context of 5G networks. Examples of such applications are industrial automation, reliable remote control, tactile internet, augmented or virtual reality (AR/VR), and autonomous driving, where the mission criticality is a common characteristic.

Depending on the application, the requirements for the end-to-end (E2E) reliability and round-trip latency might vary from $1-10^{-5}$ (i.e., 5 nines) to $1-10^{-9}$ (i.e., 9 nines) and 1 ms to 15 ms, respectively, for a packet size of 32 to 300 bytes [1]. By contrast, the typical target block error rate (BLER) in today's 4G systems using channel coding solely is $10^{-1}$, and it can be reduced to $10^{-2}$ by applying retransmission mechanisms [2]. Current latency observed in LTE networks has large variance, nevertheless, according to the measurements performed by the UK office of communications, the 4G networks round trip latency is around 53.1 ms where only 10% of the traffic experience a delay lower than 40 ms [3]. Evidently, there is a big gap between the existing performance and URLLC requirements that drive the need to re-think traditional capacity-oriented approaches by introducing guaranteed service methods.

One important challenge to tackle is on how to decrease error probability by several orders of magnitude. Multi (dual) connectivity is one of the promising tools to improve the reliability and capacity of mobile radio networks [4]. It allows users to connect to different types of base stations (BSs) or distinct carriers such that multiple copies of the same information can be delivered using different radio access technologies [5].

In URLLC context, the recent literature has studied the performance of multi-connectivity with packet duplication either to enhance reliability or to improve other network parameters such as required transmit power, link utilization and network power consumption for a specific reliability requirement (e.g., [4], [6]). The idea is to transmit multiple instances of a packet over multiple links. For instance, [6] shows that the use of packet duplication can, in certain scenarios, lead to an improvement of resource utilization in terms of the number of used physical resource blocks. However, the supporting analysis is conducted under the assumption of independent links for radio access, i.e., without shadowing cross-correlation. According to the simulation results in [6], [7], and a mathematical framework developed for device-to-device communication systems in [9], the shadowing cross-correlation between different wireless links has significant impact on the reliability. Nevertheless, analyzing the impact of correlation on the E2E network reliability and its consideration on the design of the network supporting URLLC applications is still an open problem. Moreover, the impact of a single point of failure has been neglected. The main contributions of this paper are to investigate the impact of different failure correlation factors on E2E reliability and analyze its significance on the design of a system supporting URLLC applications.

The remainder of this paper is organized as follows. We describe our system model in Section II, and discuss the causes of correlated failures in Section III. An end-to-end reliability

This work was supported by Swedish Foundation for Strategic Research (SSF) under Grant iPhD:ID17-0079.

analytical model for two different dual connectivity architectures is derived in Section IV. The impact of realistic correlation values on E2E reliability is evaluated in Section V through numerical analysis. Section VI summarizes concluding remarks.

## II. SYSTEM MODEL

In this paper, we consider the dual connectivity (DC) case; however, the work can be generalized for the case of multi-connectivity. We adopt the two proposed architectures for DC in 5G networks [10] and illustrate it in Fig. 1. In both cases, the user equipment (UE) communicates with both a master next generation NodeB (MgNB) and a secondary gNB (SgNB); the former being responsible for the basic communication and the latter enhancing the robustness and reliability by providing an additional link in the radio access network (RAN) part. In the core network (CN) part, multiple forwarding nodes, e.g., routers, switches or repeaters, connect the gNB to where the user plane function (UPF) is deployed.

In the RAN split architecture, shown in Fig. 1(a), gNBs connect to each other via a non-ideal Xn backhaul interface. Note that this interface enables the interconnection between gNBs [11]. All instances of duplicated packets are received at MgNB, for both uplink (UL) and downlink (DL) case. MgNB duplicates the first successfully received instance and forwards it. The rest of the copies are discarded at the time of arrival at MgNB. On the downside, MgNB is the single point of failure, implying that failure in the MgNB results in losing all instances of duplicated packets. In Fig. 1(b), i.e., CN split architecture, the traffic split happens in the two edges of the network, namely UE and UPF.

We note that the UE in URLLC context is not necessarily a mobile device. As a reference, we consider an industrial automation use case where the UE is a connected actuator that is either static or moves within a limited space.

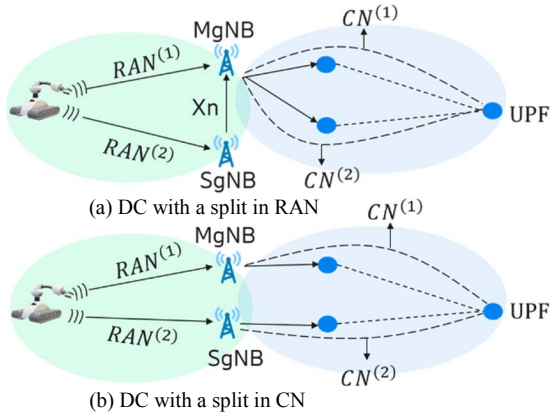

(a) DC with a split in RAN

(b) DC with a split in CN

Fig. 1. Dual connectivity architecture and split options.

When data replication is in use, the radio resource control (RRC) layer is responsible for configuring the packet duplication operation. The packet data convergence protocol (PDCP) layer duplicates the packets, and they are transmitted via separate logical legs in below layers, i.e., radio link controller (RLC), medium access control (MAC) and physical layer. In DL, packets are received via two different legs, and the PDCP layer is responsible for removing the redundant duplicated packets. In this functionality, RRC connection re-establishment or PDCP buffer overflow leads to a loss in all information copies. The former may take up to 200 ms where all the connectivity towards the UE is down.

## III. REASONS FOR CORRELATED FAILURES

The DC architectures introduced in Section II provide redundancy over parts of the end-to-end path. Measurements reveal that there exists a spatial correlation between radio links [12], which leads to a correlation in the failure events.

Failures may be correlated due to shadow fading which is caused by obstacles blocking the path between gNB and UE. Since links in geographic proximity experience common propagation area, they encounter similar environmental effects. Therefore, multiple radio links, connecting different gNBs to a given UE, experience spatially cross-correlated shadowing effects, which is usually referred as cross-correlation [13]. An example is industrial automation factories where the environment is time-varying due to constant movements of different objects, e.g., robots. When moving obstacles are in a same neighborhood of the UE, they make shadowing components correlated with each other [14]. This correlation still exists if the system uses time-frequency grid resources because the source of correlation is the physical environment and is fixed for all frequencies. For instance, the measurements in [15] show that the inter-frequency shadowing values on the same radio link can have a maximum correlation of 0.92.

Furthermore, measurement studies in [12] show that even for links toward receivers that are far apart, shadowing correlation of at least 0.05 exist. In typical cellular applications, these values can be considered negligible and therefore, the links can be assumed independent. However, due to the strict reliability requirement of up to nine 9s in URLLC, the effect of this 0.05 correlation becomes significant. Indeed, a small increase in the conditional probability of failure in the secondary link impacts the E2E reliability significantly.

## IV. ANALYTICAL MODEL

In this section, we derive expressions for the E2E reliability for the two DC architectures when correlated failures are taken into consideration. Thereafter, we define reliability, denoted by $R$, as the probability that a message is successfully transmitted from source to destination and decoded within a delay requirement. The error rate $\varepsilon = 1 - R$ is the probability that a message with a specific size is not successfully transmitted from source to destination within a delay requirement.

### A. RAN Reliability

In URLLC, the achievable reliability between UE and gNB over path $i$, $R_{RAN^{(i)}}$, can be defined as

$$R_{RAN^{(i)}} = P\big(D^{(i)} \leq D_{max}, P_r^{(i)} \geq P_{th}\big), \quad (1)$$

where $D^{(i)} \leq D_{max}$ ensures that the observed delay for a packet is lower than the defined URLLC requirement for RAN with a latency bound of $D_{max}$. Besides, $P_r^{(i)} \geq P_{th}$ ensures that the received power from the $i$-th path is higher than a minimum threshold $P_{th}$, required for achieving the URLLC requirement. Similarly, $\varepsilon_{RAN^{(i)}}$ is the error rate on the $i$-th path on the RAN part and $\varepsilon_{Xn}$ is the error rate of the Xn interface.

Let us define $I_{RAN_f^{(i)}}$ as the indicator function of $RAN_f^{(i)}$, i.e., the failure event in the $i$-th link in the RAN including the wireless link as well as RLC, MAC and physical layer. The RAN correlation, $\rho$, between the failure events on the two RAN links is

$$\rho = \rho_{RAN_f^{(1)}, RAN_f^{(2)}}$$

$$= \frac{E\left[I_{RAN_f^{(1)}} I_{RAN_f^{(2)}}\right] - E\left[I_{RAN_f^{(1)}}\right] E\left[I_{RAN_f^{(2)}}\right]}{\sigma_{I_{RAN_f^{(1)}}} \sigma_{I_{RAN_f^{(2)}}}}$$

$$= \frac{P\left(RAN_f^{(1)}, RAN_f^{(2)}\right) - \varepsilon_{RAN^{(1)}} \varepsilon_{RAN^{(2)}}}{\sigma_{I_{RAN_f^{(1)}}} \sigma_{I_{RAN_f^{(2)}}}}, \quad (2)$$

where $E$ and $\sigma$ denote the expectation and standard deviation function. Since $I_{RAN_f^{(1)}}$ and $I_{RAN_f^{(2)}}$ are Bernoulli variables, then

$$\sigma_i = \sigma_{I_{RAN_f^{(i)}}} = \sqrt{\varepsilon_{RAN^{(i)}}(1 - \varepsilon_{RAN^{(i)}})}. \quad (3)$$

As a result, we can derive the probability of a joint failure for both $RAN_f^{(1)}$ and $RAN_f^{(2)}$ as

$$P\left(RAN_f^{(1)}, RAN_f^{(2)}\right) = \varepsilon_{RAN^{(1)}} \varepsilon_{RAN^{(2)}} + \rho \sigma_1 \sigma_2. \quad (4)$$

In the literature, measurements are mostly performed to quantify shadowing cross-correlation. Assuming that the delay requirements in (1) are met on both wireless links, the event correlation represents the joint probability of shadowing to be less than a threshold on two different links. Table I reports some examples of the mapping between the shadowing correlation and failure event correlation with $\varepsilon_{RAN} = 10^{-4}$. A generalized mapping is reported in the Appendix.

TABLE I

EXAMPLES OF THE MAPPING BETWEEN SHADOWING CORRELATION, $\rho_h$, AND EVENT CORRELATION, $\rho$, FOR $\varepsilon_{RAN} = 10^{-4}$.

| $\rho_h$ | 0.05 | 0.1 | 0.3 | 0.7 | 1 |
|---|---|---|---|---|---|
| $\rho$ | 0.0001 | 0.0003 | 0.004 | 0.1 | 1 |

*B. CN Reliability*

Regarding the DC architectures in Fig. 1, the error rate of a single path in the CN, $\varepsilon_{CN^{(i)}}$, can be derived as

$$\varepsilon_{CN^{(i)}} = 1 - \left(1 - \varepsilon_{l,N+1^{(i)}}\right) \prod_{j=1}^{N} \left(1 - \varepsilon_{n,j^{(i)}}\right)\left(1 - \varepsilon_{l,j^{(i)}}\right), \quad (5)$$

where $N$ is the total number of intermediate nodes, $\varepsilon_{n,j^{(i)}}$ represent the $j$-th intermediate node error rate on path $i$ and $\varepsilon_{l,j^{(i)}}$ is the error rate of the link connected to it. Note that the definition in (5) includes neither gNBs nor UPF error rate.

Besides, the correlations between parallel and serial links in (5) are assumed negligible.

*C. E2E Reliability*

The E2E reliability is derived for both architectures in Fig. 1. Let us denote the E2E reliability by $R_{e2e,r}$ and $R_{e2e,c}$ for DC architecture with a split in RAN and CN, respectively.

*1) DC with Split in RAN*

To derive the E2E reliability of the architecture depicted in Fig. 1(a), we consider that the dual connectivity is separated on the RAN and CN, meaning that the PDCP layer in MgNB is responsible for handling the duplication by means of either duplicating before the transmission or duplication removal when receiving an instance. Therefore, the E2E reliability can be written as

$$R_{e2e,r} = (1 - \varepsilon_m)(1 - \varepsilon_{Mg})\left(1 - \prod_{i=1}^{n} \varepsilon_{CN^{(i)}}\right)(1 - \varepsilon_{UPF}), \quad (6)$$

where $\varepsilon_m$ is the observed error rate between UE and MgNB, $\varepsilon_{Mg}$ is the error rate of MgNB, $n$ is the number of CN paths to UPF and is equal to two in our model, and $\varepsilon_{UPF}$ is the UPF error rate. Note that the number of CN redundant paths depends on the E2E requirements. An improved tunneling protocol may be required to enable the parallel transmission of duplicated packets [10]. On the other hand, the E2E reliability is heavily dependent on $\varepsilon_{Mg}$ in this architecture.

Assuming independence of $RAN_f^{(2)}$ and $Xn_f$, then $\varepsilon_m$ is

$$\varepsilon_m = P(UE_f) + P(UE_s)P\left(RAN_f^{(1)}, RAN_f^{(2)}\right) \\ + P(UE_s)P\left(RAN_f^{(1)}, RAN_s^{(2)}\right)P(Sg_f \cup Xn_f). \quad (7)$$

The first joint probability on the right side of the equation is provided in (4) and the second one, as a joint probability of failure event in the first link and success event in the second one, can also be derived using (4) and considering

$$\rho_{RAN_f^{(1)}, RAN_s^{(2)}} = -\rho_{RAN_f^{(1)}, RAN_f^{(2)}}. \quad (8)$$

Moreover, $P(Sg_f)$ and $P(Xn_f)$ are the failure probability in SgNB and Xn, respectively. Besides, $P(UE_f)$ and $P(UE_s)$ are the UE failure and success probability, respectively, captured in RRC and PDCP layer. Putting all pieces together, the E2E reliability becomes

$$R_{e2e,r} = (1 - \varepsilon_{Mg})(1 - \varepsilon_{CN^{(1)}}\varepsilon_{CN^{(2)}})(1 - \varepsilon_{UPF})(1 - \varepsilon_{UE}) \\ \left(1 - \varepsilon_{RAN^{(1)}}\varepsilon_{RAN^{(2)}} - \varepsilon_{RAN^{(1)}}(1 - \varepsilon_{RAN^{(2)}})\varepsilon_{SX} \\ - \rho \sigma_1 \sigma_2 (1 - \varepsilon_{SX})\right), \quad (9)$$

where $\varepsilon_{UE}$ is the UE error rate and $\varepsilon_{SX}$ is defined as

$$\varepsilon_{SX} = P(Sg_f \cup Xn_f). \quad (10)$$

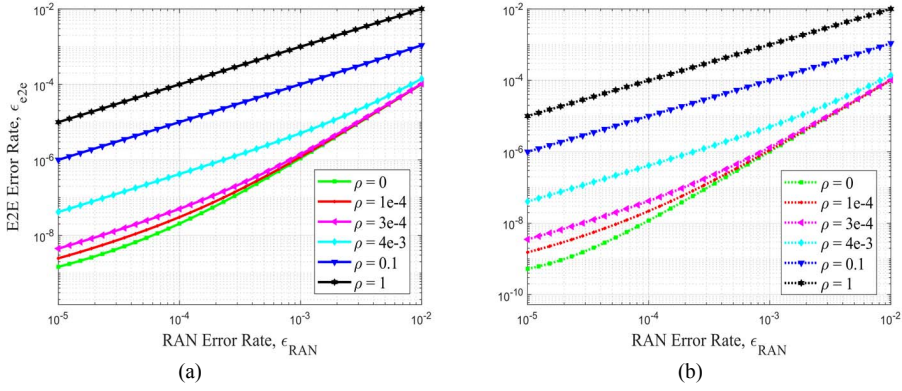
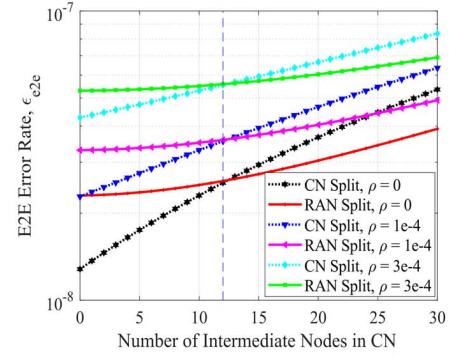

Fig. 2. The impact of correlation on E2E error rate for different RAN error rates when there is only one intermediate node in the core network for (a) RAN split architecture with $\varepsilon_{SX}$ of $10^{-4}$, and (b) core network split architecture.

Fig. 3. The E2E error rate for different numbers of intermediate nodes in the core network where $\varepsilon_{RAN} = \varepsilon_{SX} = 10^{-4}$

## *2) DC with Split in CN*

In the architecture in Fig. 1(b), the packets are duplicated in UE PDCP layer for UL and the UPF for DL. Both entities are also responsible for the removal of the duplicated instances in case another instance has already been received. Contrary to the RAN split architecture, there is no single point of failure en route in this architecture.

For notation simplicity, we can include the gNBs error rate into $\varepsilon_{CN^{(i)}}$ in (5) as

$$\varepsilon_{Cg^{(i)}} = 1 - (1 - \varepsilon_{CN^{(i)}})(1 - \varepsilon_g), \quad (11)$$

where $\varepsilon_g$ denotes the error rate of the gNB that is connected to path $i$.

Assuming the independence of CN and RAN, the E2E reliability is given by

$$R_{e2e,c} = 1 - P(UE_f \cup UPF_f)$$
$$- P(UE_s)P(UPF_s)\Big(P\big(RAN_f^{(1)}, RAN_f^{(2)}\big)$$
$$+ \varepsilon_{Cg^{(1)}}P\big(RAN_s^{(1)}, RAN_f^{(2)}\big) + \varepsilon_{Cg^{(2)}}P\big(RAN_f^{(1)}, RAN_s^{(2)}\big) \quad (12)$$
$$+ \varepsilon_{Cg^{(1)}}\varepsilon_{Cg^{(2)}}P\big(RAN_s^{(1)}, RAN_s^{(2)}\big)\Big),$$

where the joint RAN probabilities can be calculated using (4) and (8) and putting everything together, $R_{e2e,c}$ is

$$R_{e2e,c} = (1 - \varepsilon_{UE})(1 - \varepsilon_{UPF})$$
$$\Big(1 - \varepsilon_{Cg^{(1)}}\big(\varepsilon_{RAN^{(2)}}(1 - \varepsilon_{RAN^{(1)}}) - \rho\,\sigma_1\sigma_2\big)$$
$$- \varepsilon_{Cg^{(2)}}\big(\varepsilon_{RAN^{(1)}}(1 - \varepsilon_{RAN^{(2)}}) - \rho\,\sigma_1\sigma_2\big) \quad (13)$$
$$- \varepsilon_{Cg^{(1)}}\varepsilon_{Cg^{(2)}}\Big((1 - \varepsilon_{RAN^{(1)}})(1 - \varepsilon_{RAN^{(2)}})$$
$$+ \rho\,\sigma_1\sigma_2\Big) - \varepsilon_{RAN^{(1)}}\varepsilon_{RAN^{(2)}} - \rho\,\sigma_1\sigma_2\Big).$$

## V.  PERFORMANCE EVALUATION

For the presentation of the results of the analysis, without loss of generality, it is assumed that the reliability of different radio links in the RAN part is the same. Moreover, the two paths between gNB and UPF in the CN have the same number of hops. Besides, the $\varepsilon_{UE}$, $\varepsilon_g$ and $\varepsilon_{UPF}$ are assumed to be $10^{-10}$, and the $\varepsilon_{n,j^{(i)}}$ and $\varepsilon_{l,j^{(i)}}$ in (5) are assumed to be $10^{-7}$ and $4 \times 10^{-6}$, respectively, for $\forall\, i, j$, unless otherwise stated.

Fig. 2 illustrates the $\varepsilon_{e2e}$ for different values of $\varepsilon_{RAN}$ and $\rho$ where the number of intermediate nodes in the CN part is fixed to be only one. We see that the E2E error rate can change by orders of magnitude with even small values of correlation. For example, in the case of $\varepsilon_{RAN} = 10^{-4}$, the $\varepsilon_{e2e}$ increases for both architectures by more than an order of magnitude when correlation assumption changes from 0, i.e., independent links, to 0.004. Note that an event correlation of 0.004 is obtained, in the absence of additional factors, with a shadowing cross-correlation value of 0.3 (Table I), which is commonly measured in URLLC deployments [14]. In this case, the analysis based on the assumption of independent links could mislead the design in a sense that a predicted reliability of eight 9s is actually providing six 9s. We also observe, in other experiments not shown here, that the DC performance in terms of E2E error rate is lower bounded by the error rate of a single point of failure, i.e., $\varepsilon_{UE}$, $\varepsilon_{UPF}$ as well as $\varepsilon_{Mg}$ for the RAN split architecture.

The number of intermediate nodes is the parameter to determine the service distance in the CN. In Fig. 3, we show how the two architectures behave with different values of correlation when the number of intermediate nodes varies from 0 to 30. It is interesting to observe that there is a trade-off between the two design options. The superior architecture in terms of reliability depends on the choice of service distance. The further we deploy the UPF, the reliability provided by CN decreases and hence, the design which provides a full diversity in core network becomes more effective. In this case, RAN split architecture which ensures the duplication at MgNB can compensate for the higher service distance. In Fig. 4, we derive the optimal architecture in terms of reliability for different system parameters. The green area shows where the reliability of RAN split architecture is higher than the CN split architecture and yellow is vice-versa. For shorter service distances, where the CN is more robust, the Xn interface is the limiting factor. This leads to better

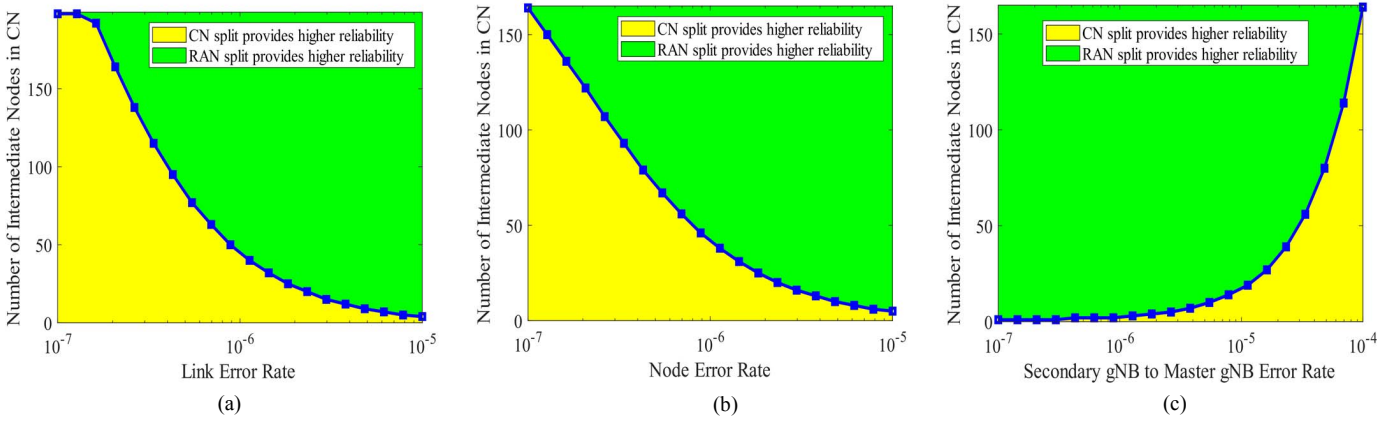

Fig. 4. The optimal architecture in terms of E2E reliability for different number of intermediate nodes in the CN, where green area shows where the reliability of RAN split architecture is higher than CN split architecture and yellow is vice-versa. In (a) $\varepsilon_{SX}$ is $10^{-4}$, and the default node error rate of $10^{-7}$ is assumed. In (b) the default link error rate of $4 \times 10^{-6}$ and $\varepsilon_{SX} = 10^{-4}$, and in (c) the default link error rate and node error rate is assumed.

performance on CN split architecture where we have E2E duplication and no Xn interface is required.

Accordingly, the presence of correlation can influence the choice of the architecture to fulfill the requirement on E2E error rate. This is illustrated in Fig. 5 where the maximum number of intermediate nodes that can fulfill the E2E error rate requirement of $3 \times 10^{-8}$ is calculated over different values of link error rate. The two architectures are analyzed assuming independent and correlated RAN links. As can be seen, under the assumption of independent RAN links, the RAN split option can fulfill the E2E reliability requirement with higher numbers of intermediate nodes, i.e., bigger service distance, compared to CN split option. On the other hand, the CN split option outperforms the RAN split architecture when there exists only a small correlation between RAN links. For example, for $\varepsilon_{l,j^{(i)}}$ of $10^{-6}$ for $\forall i, j$, the RAN split architecture cannot fulfill the requirement even with a direct link between gNB to UPF while the system can be designed with a maximum of 36 intermediate nodes with CN split architecture.

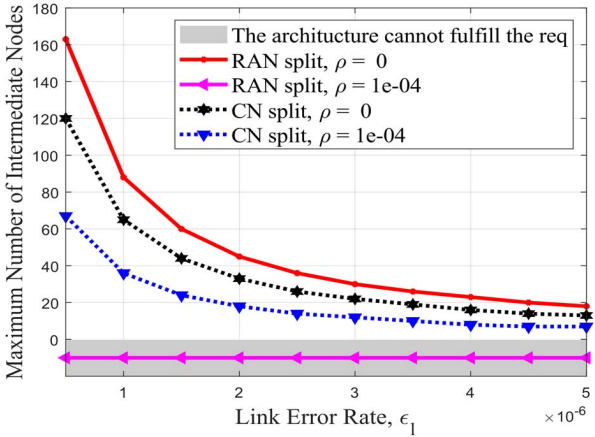

Fig. 5. The maximum number of intermediate nodes for different values of $\varepsilon_l$ in CN when E2E error rate requirement is $3 \times 10^{-8}$ and $\varepsilon_{RAN} = \varepsilon_{SX} = 10^{-4}$

## VI. CONCLUSIONS

In this paper, we investigated the impact of correlation on E2E reliability in the context of URLLC. We developed an E2E analytical model which incorporates correlated failure scenarios to analyze packet duplication technique in DC architectures. Our results show that small values of correlation coefficients lead to several orders of magnitude impact on the E2E error rate and, more importantly, lead to a completely different outcome for the alternative architectures in terms of feasibility and service distance for a given E2E requirement. By including shadowing, our model enables more realistic design of a robust DC architecture. As future work, the analytical model may be extended to generalize the DC case to multi-connectivity architectures.

### ACKNOWLEDGMENT

P. Di Marco acknowledges the support by the Italian Government under the PON Ricerca & Innovazione 2014-2020 (AIM 1877124).

## APPENDIX

We defined the radio access network (RAN) reliability, $R_{RAN^{(i)}}$, as the probability of success for sending a packet from the UE to the $i$-th gNB in (1).

In the following, we derive the failure event correlation, assuming that shadowing correlation is the only reason for having correlated events and the delay requirements in (1) are met on both wireless links. This implies that the delay components (i.e., processing, transmission, propagation, and queuing delay) are assumed bounded and we consider a fixed number (or no) retransmissions so that the delay is lower than the $D_{max}$ with probability 1, independently of the success rate.

On a particular wireless link, the measured received power can be modeled as

$$P_r^{(i)} = P_t^{(i)} - P_l^{(i)} - X_{dB,i}, \quad (14)$$

where $P_r^{(i)}$, $P_t^{(i)}$ and $P_l^{(i)}$ are the received power, transmit power and path loss (all in dB) on the $i$-th path, respectively. Besides, $X_{dB,i}$ is a Gauss-distributed random variable with mean 0 and variance $\sigma_{dB,i}^2$ representing the variations in received power on the $i$-th RAN due to shadow fading [16]. Therefore, the RAN reliability, in (1), is given by

$$R_{RAN^{(i)}} = P\big(P_{th}^{(i)} < P_t^{(i)} - P_l^{(i)} - X_{dB,i}\big) \\ = P\left(\frac{X_{dB,i}}{\sigma_{dB,i}} < \frac{P_t^{(i)} - P_l^{(i)} - P_{th}^{(i)}}{\sigma_{dB,i}}\right). \quad (15)$$

Let us define the mean normalized received power above threshold for the $i$-th path in RAN, $\beta_i$, as

$$\beta_i = \frac{P_t^{(i)} - P_l^{(i)} - P_{th}^{(i)}}{\sigma_{dB,i}}, \quad (16)$$

therefore, using (3) the RAN reliability becomes

$$R_{RAN^{(i)}} = 1 - Q(\beta_i), \quad (17)$$

where $Q(\beta_i)$ is the Q-function of the mean normalized received power above threshold.

The failure event correlation among two different wireless links, $\rho$, can be derived using bivariate Gaussian distribution as

$$\rho = \frac{P(X_{dB,1} > \beta_1, X_{dB,2} > \beta_2) - \varepsilon_{RAN^{(1)}}\varepsilon_{RAN^{(2)}}}{\sigma_{I_{RAN_f^{(1)}}}\sigma_{I_{RAN_f^{(2)}}}}$$

$$= \frac{\int_{\beta_2}^{\infty}\int_{\beta_1}^{\infty}\frac{1}{2\pi\sqrt{1-\rho_h^2}}e^{\frac{-(X_{dB,1}^2 + X_{dB,2}^2 - 2\rho_h X_{dB,1} X_{dB,2})}{2(1-\rho_h^2)}}dX_{dB,1}\,dX_{dB,2}}{\sigma_{I_{RAN_f^{(1)}}}\sigma_{I_{RAN_f^{(2)}}}}$$

$$- \frac{Q(\beta_1)Q(\beta_2)}{\sigma_{I_{RAN_f^{(1)}}}\sigma_{I_{RAN_f^{(2)}}}}$$

$$= \frac{\frac{1}{\sqrt{2\pi}}\int_{\beta_2}^{\infty}Q\left(\frac{\beta_1 - \rho_h X_{dB,2}}{\sqrt{1-\rho_h^2}}\right)e^{\frac{-X_{dB,2}^2}{2}}dX_{dB,2} - Q(\beta_1)Q(\beta_2)}{\sigma_{I_{RAN_f^{(1)}}}\sigma_{I_{RAN_f^{(2)}}}}, \quad (18)$$

where $\rho_h$ represents the shadowing cross-correlation between the two RAN links.